# AT VOYAGER 1 STARTING ON ABOUT AUGUST 25, 2012 AT A DISTANCE OF 121.7 AU FROM THE SUN, A SUDDEN SUSTAINED DISAPPEARANCE OF ANOMALOUS COSMIC RAYS AND AN UNUSUALLY LARGE SUDDEN SUSTAINED INCREASE OF GALACTIC COSMIC RAY H AND HE NUCLEI AND ELECTRONS OCCURRED

W.R. Webber[1], F.B. McDonald[2+], A.C. Cummings[3], E.C. Stone[3],

B. Heikkila[5] and N. Lal[5]

1.  New Mexico State University, Department of Astronomy, Las Cruces, NM  88003, USA

2. University of Maryland, Institute of Physical Science and Technology, College Park, MD  20742, USA

3.  California Institute of Technology, Space Radiation Laboratory, Pasadena, CA  91125, USA

4.  NASA/Goddard Space Flight Center, Greenbelt, MD  20771, USA

+    Deceased August 31, 2012

**ABSTRACT**

At the Voyager 1 spacecraft in the outer heliosphere, after a series of complex intensity changes starting at about May 8th, the intensities of both anomalous cosmic rays (ACR) and galactic cosmic rays (GCR) changed suddenly and decisively on August 25th (121.7 AU from the Sun). The ACR started the intensity decrease with an initial e-folding rate of intensity decrease of ~1 day. Within a matter of a few days, the intensity of 1.9-2.7 MeV protons and helium nuclei had decreased to less than 0.1 of their previous value and after a few weeks, corresponding to the outward movement of V1 by ~0.1 AU, these intensities had decreased by factors of at least 300-500 and are now lower than most estimates of the GCR spectrum for these lower energies and also at higher energies. The decrease was accompanied by large rigidity dependent anisotropies in addition to the extraordinary rapidity of the intensity changes. Also on August 25th the GCR protons, helium and heavier nuclei as well as electrons increased suddenly with the intensities of electrons reaching levels ~30-50% higher than observed just one day earlier. This increase for GCR occurred over ~1 day for the lowest rigidity electrons, and several days for the higher rigidity nuclei of rigidity ~0.5-1.0 GV. After reaching these higher levels the intensities of the GCR of all energies from 2 to 400 MeV have remained essentially constant with intensity levels and spectra that may represent the local GCR. These intensity changes will be presented in more detail in this, and future articles, as this story unfolds.

## **Introduction**

The passage of the Voyagers 1 and 2 spacecraft through the outer heliosphere (heliosheath) has revealed a region quite unlike the inner heliosphere inside the heliospheric termination shock (HTS). The radial solar wind speed slows down from ~400 km/s to ~130 km/s (Richardson, et al., 2008) and later at about 20 AU beyond the HTS may decrease to very low values (Krimigis, et al., 2011). The anomalous and galactic cosmic ray (ACR and GCR) intensities hardly changed at the HTS contrary to theoretical expectations (Stone, et al., 2005). The magnetic field shows many distinct structures or features associated with the HTS and the heliosheath region beyond (Burlaga and Ness, 2010). One of the largest of these structures was encountered by V1 at 2009.7 when the spacecraft was ~17 AU beyond the HTS crossing distance of 94 AU. At this time the field direction suddenly changed from 90° to 270° possibly indicating a sector crossing. Also at this time the galactic cosmic ray electron intensity increased by an unprecedented 30% and the radial intensity gradients of these electrons and higher energy nuclei decreased by over a factor of two (Webber, et al., 2011). Sudden intensity increases of electrons and nuclei again occurred about 1.5 years later at a distance of ~117 AU or 23 AU beyond the HTS crossing distance (Webber, et al., 2011). More recently a new series of changes have been observed starting at about 2012.0 in both GCR and ACR. In particular at about 2012.35 at ~120.5 AU from the Sun, large increases of both GCR nuclei and electrons were observed with little corresponding changes of ACR. In fact, throughout most of these unusual GCR intensity changes in the outer heliosphere, the ACR H, He and O nuclei from ~1-50 MeV also changed, but only modestly, and the changes in ACR and GCR were not always correlated.

The ACR represent the dominant energetic population in the heliosheath above ~1 MeV with intensities ~$10^2$-$10^3$ times those observed in the heliosphere inside the HTS. These ACR particles are accelerated somewhere in the heliosheath (several mechanisms are possible) and remain quasi trapped there, leaking into the inner heliosphere where they are only weakly observed at the Earth. At the outer boundary of the heliosheath these particles may also leak out into the interstellar region.

On August 25[th] when V1 was at 121.7 AU from the Sun the intensity of the ACR component began to decrease rapidly. Within a few days the intensity of this dominant energetic component above 1-2 MeV decreased by more than 90-95% reaching intensity levels not seen at V1 since it was well inside the HTS. At the same time a sudden increase of a factor of ~2 occurred in lower energy (6-100 MeV) electrons and ~30-50% for the higher energy nuclei above 100 MeV. The magnitude of this intensity change for ACR has never previously been observed in the 35 year V1 mission except for the Jupiter encounter. This simultaneous reduction of ACR intensities at lower energies and the abrupt increase in GCR intensities at higher energies has suddenly revealed one of the holy grails of GCR studies, the actual local interstellar spectra (LIS) of the GCR nuclei from H to Fe above ~10-20 MeV and possibly even to lower energies. For the multi-dimensional CRS

88  instrument used here (Stone, et al., 1977), the intrinsic backgrounds are so low that the reduction of
89  ACR is at least a factor ~300-500 making the low energy GCR measurements possible.

90      This large decrease of ACR was preceded by 2 precursor temporary decreases starting on
91  July 28[th] and August 14[th]. The decreases on July 28[th] and August 14th may indicate that the
92  boundary that has been crossed on August 25[th] is moving outward at nearly the V1 speed but that
93  this motion was irregular. Thus V1 may have crossed a boundary, which itself was very sharp, at
94  least 5 times during this time period.

95      It is this transition into a new region and some of its implications that we wish to summarize
96  in this paper. Further details of these remarkable events will be presented in subsequent articles.

97  **The Data**

98      In Figure 1 we show the time history from 2009 to the present of GCR nuclei >200 MeV
99  and electrons of 6-14 MeV as well as ~1 MeV H nuclei as representative of the lower energy
100 energetic particle population. The complexity of the intensity changes in this time period,
101 particularly for the GCR nuclei and electrons is evident. In Figure 2 we zoom in on the time period
102 from 2012.2 to 2012.8 for the same particles. The suddenness and decisiveness of the last change in
103 the sequence at a time of 2012.65 (August 25[th]) is extraordinary. The increase at 2012.35 which
104 applied to GCR particles but not to the lowest energy ACR and also the two precursor changes at
105 2012.57 and 2012.615 which applied to both GCR and ACR are also significant and will be
106 discussed later. In the two precursor intensity changes the ACR decreased considerably and the
107 GCR increased, mimicking the last change at 2012.65 seen in the figure.

108     In the time period of just a few weeks after 2012.65 (1 week = 0.07 AU of outward
109 movement by V1) the ACR intensity between 2-10 MeV dropped to levels of ~0.2-0.3 of one
110 percent or less of the values seen earlier at these energies and the GCR nuclei and electron
111 intensities increased by ~1.5 times and 2.0 times respectively, to the highest levels observed since
112 the launch of V1. These GCR intensities have remained almost constant at these highest levels for
113 several months.

114     To see how these intensity changes affect the energy spectra of protons and helium nuclei
115 being measured at V1 we show Figures 3 and 4. For protons in Figure 3 note that the CRS particle
116 telescopes measure the proton spectra from ~2-400 MeV, for helium nuclei in Figure 4 the energy
117 range is ~2-600 MeV. Over the entire energy range the CRIS telescopes use a 2 or 3 parameter
118 analysis (Stone, et al., 1977), which practically eliminates the background that is inherent in a single
119 parameter threshold types of analysis. This enables us to accurately measure these low intensities

120     It is seen that for both protons and helium nuclei the large reduction in ACR intensity at
121 lower energies produces intensity levels only seen previously in the inner heliosphere at quiet times.
122 The intensities of both H and He nuclei that are remaining after about 2012.75 look much like some
123 earlier predictions of possible GCR spectra, perhaps down to ~10 MeV and below. So the

124 "boundary" that V1 has just crossed is an extremely effective barrier for the ACR, reducing the
125 intensities by >99% in less than 0.2 AU. It is also an effective barrier for GCR flowing inward.

126      As for the spectra that are observed for GCR protons and He after about 2012.75, they will
127 be discussed next in this paper and more fully in subsequent papers. It should be noted that this is,
128 as, yet, only a brief glimpse of these GCR and ACR particles at a location still in close proximity to
129 the boundary just crossed. Further surprises may be in store as V1 proceeds outward.

130 **Discussion and Summary and Conclusions**

131      The sudden and large decreases seen in all energetic ACR type particles above ~0.5 MeV at
132 2012.65 is the most dominant feature seen for these particles at V1 since its launch 35 years ago.
133 As this decrease develops it turns out that all ACR particles in the 2-50 MeV range exhibit a rigidity
134 dependent fractional intensity decrease. This intensity decrease, its suddenness and associated
135 anisotropies for nuclei above ~1 MeV define the features of the "boundary" just crossed. This is
136 illustrated in Figure 5 which shows an expanded intensity vs. time curve for several low energy
137 telescopes. In this figure the red curve is the >0.5 MeV integral intensity and the other curves are
138 the intensities from four ~1.9-2.7 MeV telescopes pointing in different directions. This shows the
139 considerable directional anisotropies present during the decrease.

140      The possible multiple boundary crossings are labeled 1-5, with the crossing 1, 3 and 5
141 representing crossings in which V1 moves from inside to outside. Crossings 1 and 5 are particularly
142 interesting. V1 appears to have 1[st] crossed the boundary during the tracking period on July 28[th].
143 The intensity >0.5 MeV decreases rapidly at a rate ~5% per hour. On the following day the
144 intensity bottoms out at a value ~40% of that in the middle of the previous day. During decrease 5
145 the intensity also drops to ~30% of its initial value in just one day. For a stationary boundary these
146 decreases would correspond to an intensity decrease e-folding distance ~0.01 AU.

147      It is also possible to think of V1 passing through a stationary medium. In this case the
148 decreases/increases would be caused by the passage of V1 through ribbons of field connected to the
149 region beyond the barrier. These ribbons would be ~0.01 and ~0.02 AU thick, respectively,
150 corresponding to the ~3 day event starting on July 28[th] and the 6-7 day event starting on August
151 14[th].

152      During the month between these two precursor decreases and the "final" decrease V1 moves
153 outward 0.3 AU. Since, in the non-stationary scenario, V1 moves from inside to outside the barrier
154 on crossings 1, 3 and 5, the barrier itself must be moving. The intensity decreases on July 28[th] and
155 August 13[th] could be the result of outward pulses of the barrier location and the intensity time
156 profiles of these decreases could then be the result of speed variations of the barrier movements.
157 The times between decreases 1, 3 and 5 which have remarkably similar initial intensity time
158 profiles, are 16 and 12 days, respectively, about 0.5 of a solar rotation period. So the precursor decreases could be the result of the boundary movement rather than V1 moving through stationary structures just inside the barrier itself.

Future study of the variations of the remaining particles below ~10 MeV and magnetic field data will help to determine whether this component is really a part of a low energy galactic component, or whether it is a weak "halo" of ACR around the heliosheath region. In any case there are good possibilities for determining the local diffusion coefficient in the region just by simply studying the behavior of the H, He and O intensities and anisotropies as V1 moves further beyond the barrier.

We should note that both the electron spectrum and the spectra of heavier nuclei from Be to Fe and including C and O nuclei are also being measured at V1 although this data is not reported here. This data can be used to compare with galactic propagated spectra and understand better the propagation and source characteristics of these never before studied low energy GCR. This includes the radioactive 10Be and other singly ionized components such as N, O, Ne and Ar that are part of the ACR component in the heliosheath, and the secondary components, B, F, etc., that are created only during propagation in the galaxy.

The electron spectra from ~2-100 MeV that V1 measures will be particularly valuable for understanding the propagation characteristics of the lowest rigidity particles in the galaxy. The electrons have the advantage that their spectrum in the galaxy below ~1 GeV can be deduced from the galactic radio synchrotron spectrum observed between 1-100 MHz. In fact, it will now be possible to obtain a consistency check between the Voyager measured electron spectrum and that predicted from the galactic radio synchrotron spectrum. This comparison utilizes the average galactic plasma parameters ($n_e$, T) and the magnetic field, B, which is also measured by Voyager.

The observed intensities of galactic H and He nuclei after about 2012.75 appear to peak at energies between ~30-100 MeV and then to decrease slightly at lower energies similar to that described in some galactic propagation models (See e.g., Putze, Maurin and Donato, 2010). There is no evidence of even a small residual solar modulation which would rapidly decrease the intensity of the lowest energy H and He nuclei. So from this it appears that V1 has exited the main solar modulation region, revealing H and He spectra characteristic of those to be expected in the local interstellar medium.

And finally what about the "boundary" or perhaps "discontinuity" is a better word, that Voyager crossed on August 25[th]? Well, it constitutes an almost impenetrable barrier for energetic heliospheric ACR nuclei accelerated and confined within. It is also by far the most significant barrier of the many barriers to the GCR particles that have been encountered by V1 in the outer heliosheath for both energetic GCR nuclei and low energy GCR electrons, both of which are moving inward. The GCR increase which occurred earlier on May 8[th] may be part of this outermost feature.

It also appears that V1 has exited the region of significant solar modulation during this time period. If the GCR and ACR intensities continue to remain at their present levels, then indeed this was a definitive boundary that resembles in many respects a "classical" heliopause, perhaps an even more impressive barrier to inward and outward transport of energetic particles that would have been expected. Future observations at V1 will hopefully settle this issue.

**Acknowledgements:** This article was conceived by our Voyager colleague, Frank McDonald, who is no longer with us. Frank, we have been working together for over 55 years to reach the goal of actually observing the interstellar spectra of cosmic rays, possibly now achieved almost on the day of your passing. You wanted so badly to finish this article that you had already started. Together we did it. Bon Voyage!

# Figure Captions

**Figure 1:** 5 day running average intensities of 0.5 MeV protons (mainly ACR), 6-14 MeV GCR electrons and >100 MeV protons from 2009.0 to the end of data. The various intensity jumps and radial gradient changes (mainly for electrons) are discussed in the text. The regions demarcated by arrows at the bottom of the figure indicate possible zones in the outer heliosheath.

**Figure 2:** Same data as in Figure 1 but on an expanded time scale from 2011.8 to 2013.0. Note the various relative intensity changes of GCR and ACR which are described in the text.

**Figure 3:** Weekly average of proton spectra from 2-400 MeV measured at V1 during the rapid decrease period starting from about 2012.61 to the end of data (each week = 0.07 AU of V1 outward movements). The reference time period is from 2011.8 to 2012.2. The ratio of intensities in percent between the final time period to those in the reference period are shown along the bottom of the graph below the average energy of each energy interval. The increasing ratio (percentage) above 10 MeV indicates the increasing contribution of GCR to the

227    total intensity in that interval.  The numbers to the left of each weekly average spectrum are the
228    weeks of the year 2012, with 41-44 in red representing a 4 week average as the intensities
229    stabilize.  Other features of the figure are discussed in the text.

230    **Figure 4:**  Same data as in Figure 3 except this is He data from 2-600 MeV/nuc.

231    **Figure 5:**  Hourly averages of various LET telescope rates each with energies from ~0.5-2.0 MeV
232    during the time period from July 24[th] to September 18[th], 2012.  Times of barrier crossings are
233    labeled 1-5.  This figure emphasizes the extraordinary suddenness of the decrease, the
234    anisotropies and the sustained intensity drop for the final decrease.

235

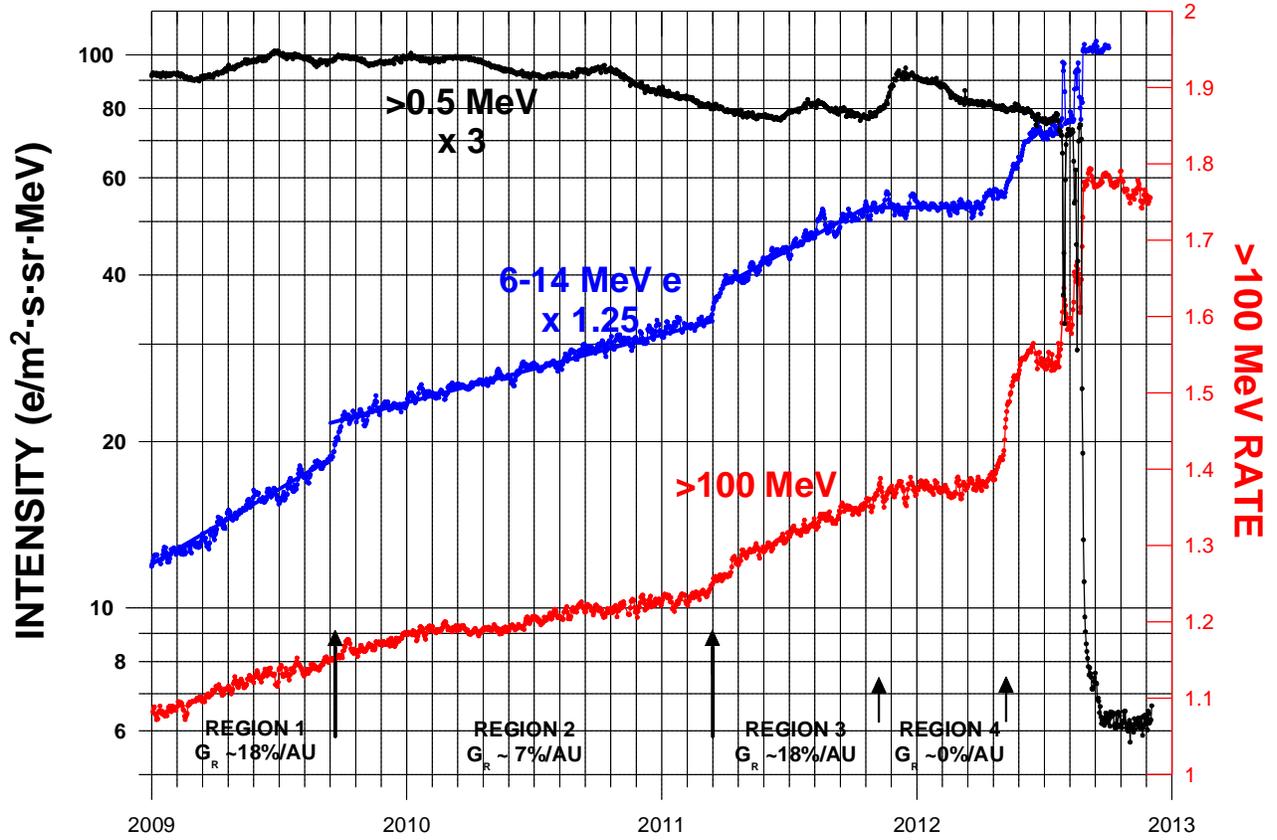

236

**FIGURE 1**

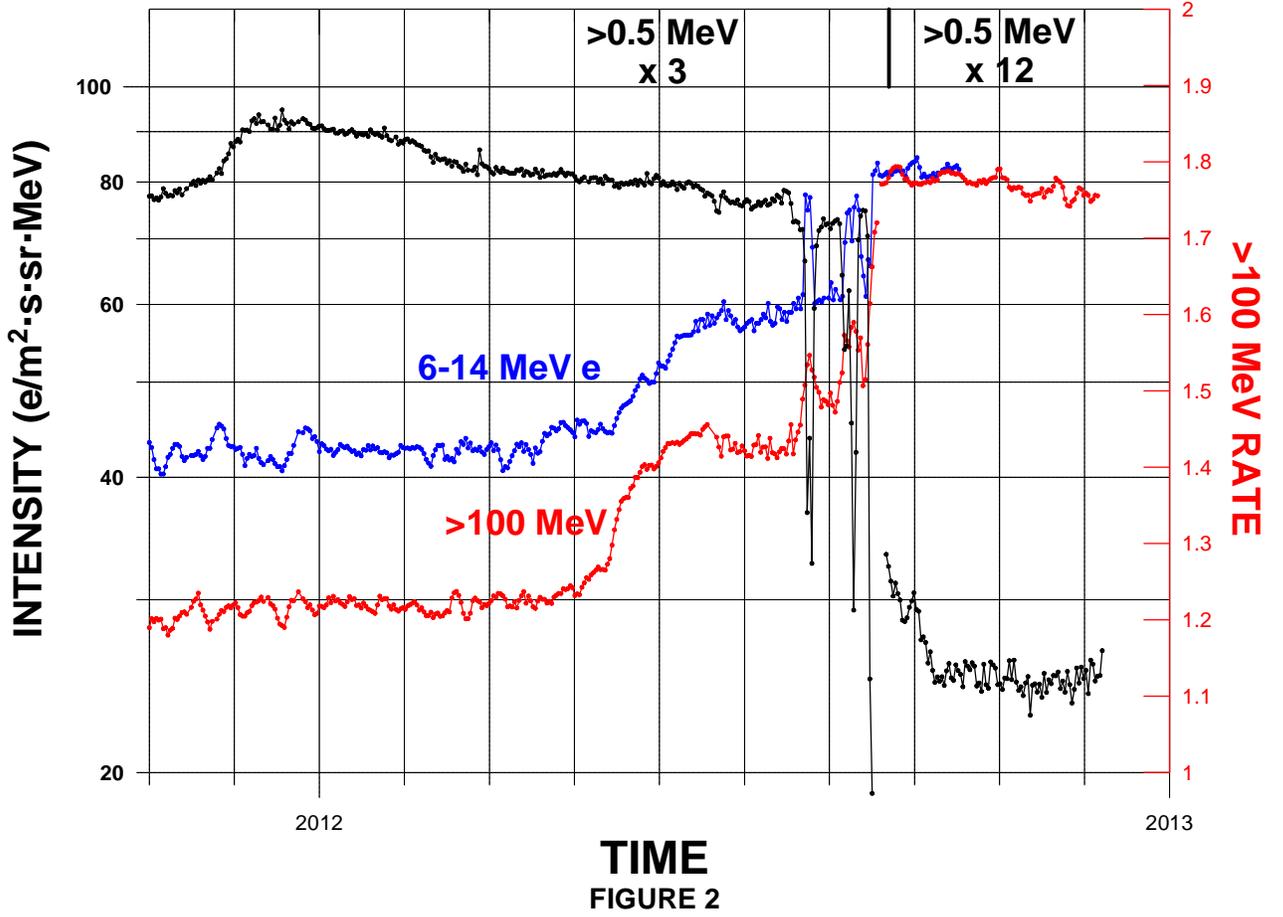

**TIME**

FIGURE 2

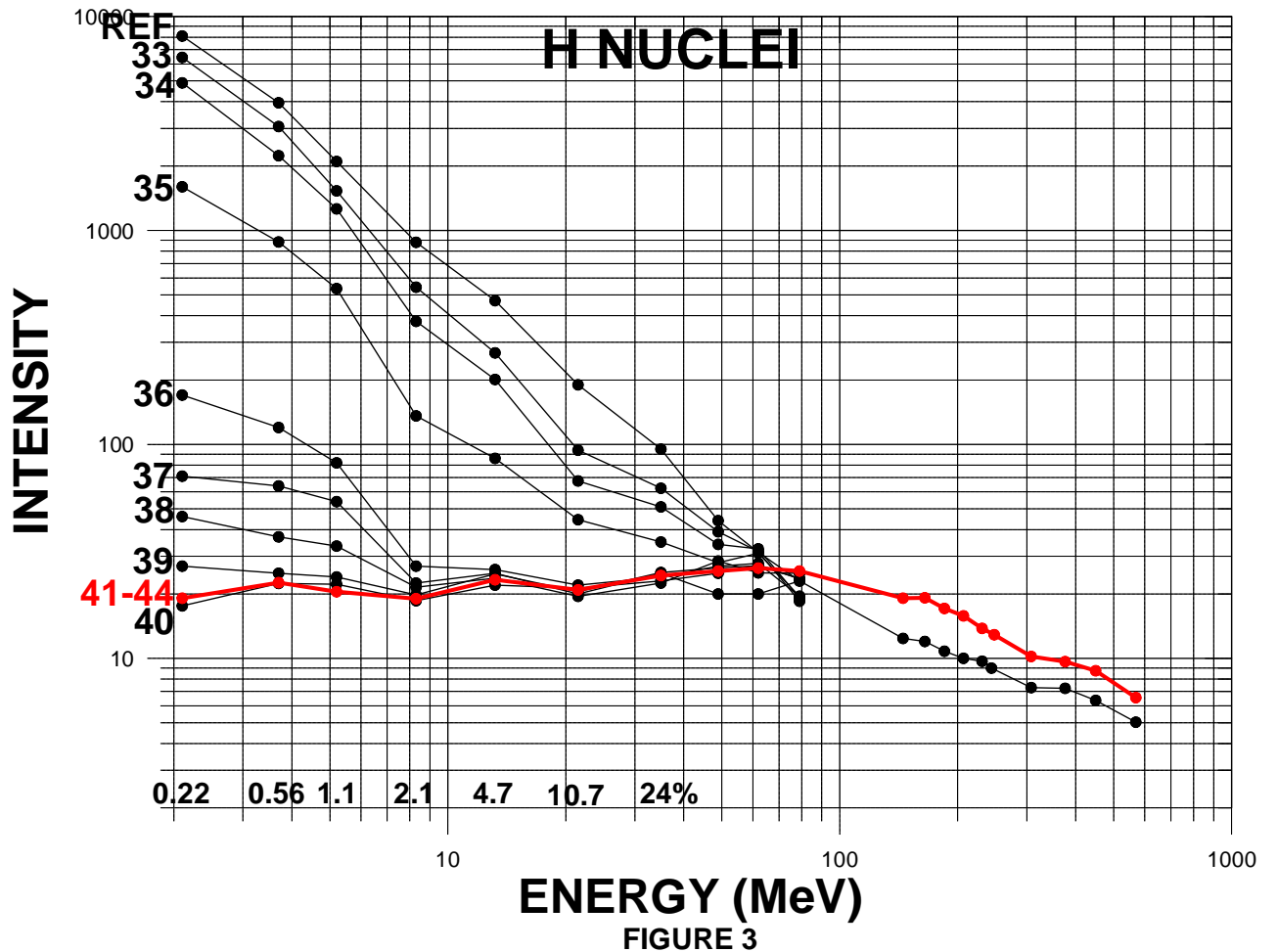

FIGURE 3

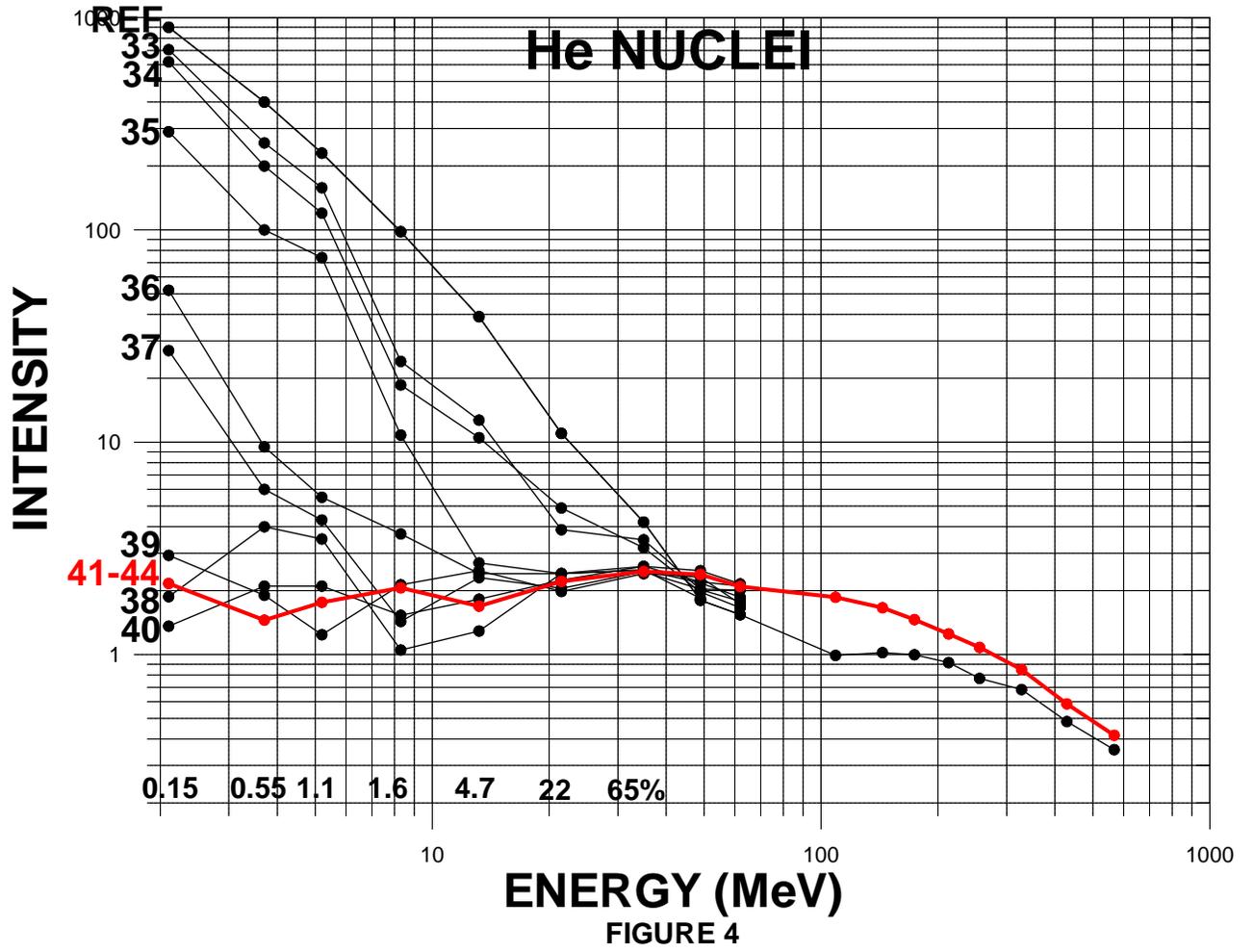

He NUCLEI

INTENSITY

ENERGY (MeV)

FIGURE 4

239

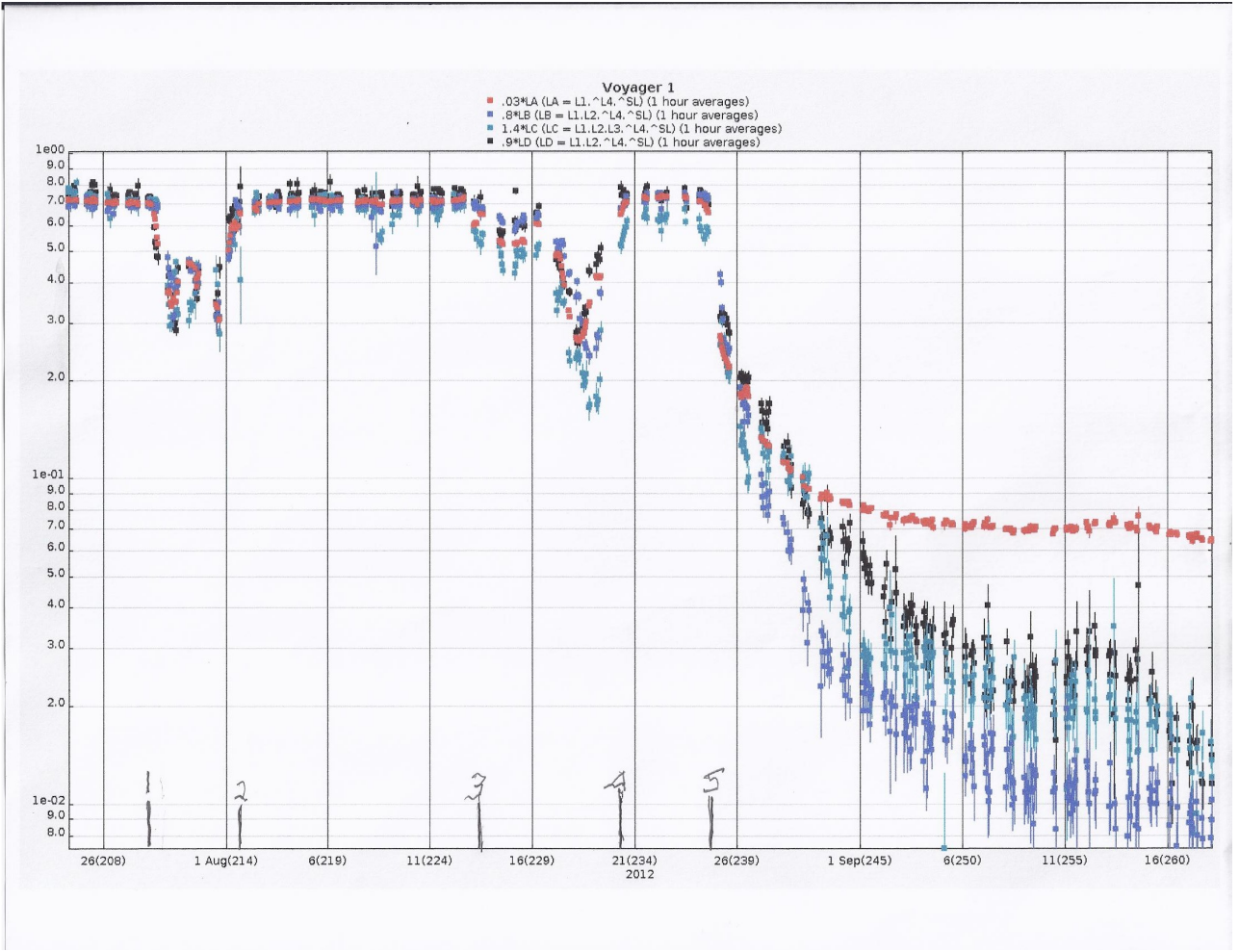

**FIGURE 5**